\documentclass[sn-mathphys-num]{sn-jnl}
\usepackage{amsmath, amssymb, amsfonts, amsthm}
\usepackage{graphicx}
\usepackage{pdflscape}
\usepackage{booktabs}
\usepackage{array}
\usepackage{xcolor}
\usepackage{textcomp}
\usepackage{multirow}
\usepackage{enumitem}
\usepackage{makecell}
\usepackage{xurl}
\usepackage{hyperref}
\usepackage{placeins}
\usepackage[title]{appendix}

\hypersetup{colorlinks=true, linkcolor=blue, citecolor=blue, urlcolor=blue}
\theoremstyle{definition}

\counterwithout{table}{section}
\counterwithout{figure}{section}
\graphicspath{{./figures/}{./}}
\newcolumntype{L}[1]{>{\raggedright\arraybackslash}p{#1}}
\newcolumntype{C}[1]{>{\centering\arraybackslash}p{#1}}

\begin{document}

% ============================================================
% Title and Author Information
% ============================================================

\title[Hospice Quality and Caregiver Experience]
{Geographic Disparities in Hospice Quality and Family Caregiver Experience: Ownership, Social Vulnerability, and County Operating Context}

\author*[1]{\fnm{Jinho} \sur{Cha}}\email{jcha@gwinnetttech.edu}
\author[2]{\fnm{Misun} \sur{Roh}}

\affil*[1]{\orgdiv{Department of Mathematics}, \orgname{Gwinnett Technical College}, \orgaddress{\state{Georgia}, \country{USA}}}
\affil[2]{\orgdiv{Clinical Nursing}, \orgname{MeSun Hospice}, \orgaddress{\city{Lawrenceville}, \state{Georgia}, \country{USA}}}

% ============================================================
% Abstract & Keywords
% ============================================================

\abstract{
Hospice quality should be interpreted in relation to both provider
organization and the local conditions under which care is delivered.
This study develops a provider--county performance assessment framework
by linking national Centers for Medicare \& Medicaid Services (CMS)
hospice data and Consumer Assessment of Healthcare Providers and Systems
(CAHPS) Hospice Survey outcomes with county measures of rurality, social
vulnerability, health burden, and workforce and health-resource context.
The adjusted analysis includes 2,928 providers in 1,078 counties and
combines geographic mapping, blockwise regression, six secondary CAHPS
outcomes, and eight sensitivity analyses.
Adding county context increased adjusted \(R^2\) from 0.077 to 0.202.
After full adjustment, for-profit hospices had overall caregiver ratings
3.832 percentage points lower than nonprofit hospices. This negative
association appeared across all six secondary CAHPS domains and remained
significant in every sensitivity specification. Higher county social
vulnerability was also associated with poorer caregiver experience,
although its magnitude depended partly on the specification of community
health burden.
These findings show that county context materially improves hospice
performance assessment but does not eliminate the ownership difference.
The framework supports context-aware monitoring, peer comparison, and
targeted quality improvement.
}


\maketitle
\vspace{0.22cm}

\clearpage
% ============================================================
% Highlights
% ============================================================
\section*{Highlights}
\begin{itemize}[leftmargin=1.5em]
\item We develop a context-aware provider--county performance assessment framework for U.S. hospice caregiver experience.
\item We link national provider-level Centers for Medicare \& Medicaid Services (CMS) files and Consumer Assessment of Healthcare Providers and Systems (CAHPS) Hospice Survey data with county-level rurality, social vulnerability, health-burden, and workforce/resource measures.
\item We interpret CAHPS caregiver-experience domains as management outcomes shaped jointly by provider organization and local operating context.
\item We combine geographic mapping, blockwise adjusted regression, six secondary outcome models, and sensitivity analyses.
\item Results show consistent negative adjusted associations for for-profit ownership and county social vulnerability, supporting targeted hospice quality-improvement monitoring.
\end{itemize}

\clearpage
% ============================================================
\section{Introduction}
\label{sec:introduction}
% ============================================================

In 2026, hospice care sits at the intersection of three increasingly
consequential questions: how the United States safeguards the quality of
end-of-life care, how Medicare resources are stewarded, and how a rapidly
changing provider sector should be held accountable. Hospice is no longer
a peripheral service. Using the latest complete national data from 2024,
the Medicare Payment Advisory Commission (MedPAC) reported that 1.82 million
Medicare beneficiaries received hospice services from approximately 6,700
providers, generating \$28.3 billion in Medicare expenditures. Hospice use
reached 52.9\% of Medicare decedents, while the number of users, total days
of care, and annual spending increased by 4.6\%, 7.7\%, and 10.4\%,
respectively \cite{MedPAC2026Hospice}. The scale and pace of this growth
make hospice performance a national management concern, not a niche issue
in palliative care.

The consequences of poor performance are unusually difficult to reverse.
Hospice shapes whether pain and other symptoms are addressed, whether help
arrives when families need it, whether communication is clear, whether
caregivers are prepared, and whether patients are treated with dignity.
These are not ordinary service encounters. Patients near the end of life
often have limited time, diminished ability to compare alternatives, and
few realistic opportunities to change providers after a failure occurs.
Family caregiver experience therefore captures more than satisfaction. It
reveals whether organizational capacity is translated into timely,
coordinated, and humane care at a point when delay, confusion, or neglect
may be irreparable. The Consumer Assessment of Healthcare Providers and Systems (CAHPS) Hospice Survey makes these experiences
measurable through nationally standardized domains covering communication,
timely help, symptom support, respect, caregiver training, overall rating,
and willingness to recommend \cite{AHRQ2026CAHPSHospice}.

Yet the national measurement system remains incomplete. Beginning in
October 2025, hospices started reporting patient-level information through
the Hospice Outcomes and Patient Evaluation (HOPE) instrument, marking an
important transition toward more clinically informative quality
measurement. In its March 2026 assessment, MedPAC concluded that an
established admission-process composite had reached ceiling levels that
limited meaningful differentiation across providers. Publicly reported
provider-level CAHPS results were also available for only slightly more
than 3,100 hospices, fewer than half of all providers
\cite{MedPAC2026Hospice}. Stable national averages may therefore coexist
with consequential variation across individual agencies, while providers
with limited public reporting remain less visible to patients, managers,
and regulators.

This measurement challenge has intensified as hospice ownership has become
more complex and potentially more consequential. Aldridge et al.
documented increasing private-equity acquisition of U.S. hospices and
showed that conventional public records do not consistently reveal the
ultimate ownership structures behind individual agencies
\cite{Aldridge2024HospicePEOpacity}. Soltoff et al. subsequently found
lower caregiver-reported quality among hospices owned by private-equity
firms or publicly traded companies than among nonprofit and other
for-profit agencies, even after adjustment for measured hospice and
community characteristics
\cite{Soltoff2024JAMAHospiceOwnership}. At the same time, recent causal
evidence indicates that for-profit hospice entry may expand access for
some beneficiaries with Alzheimer's disease and related dementias and
substitute hospice for more costly forms of care
\cite{Gruber2025DyingOrLying}. Ownership may therefore influence access,
expenditure, service organization, and family experience through distinct
and potentially competing pathways. It should be examined rather than
presumed to be uniformly beneficial or harmful.

Current federal oversight further raises the stakes. In June 2026, the
U.S. Department of Health and Human Services Office of Inspector General
reported substantial eligibility and documentation deficiencies within a
targeted sample of potentially high-risk new hospice enrollment periods.
The agency estimated that strengthened review procedures could have saved
Medicare approximately \$255.1 million within the audit sampling frame,
and the Centers for Medicare \& Medicaid Services (CMS) concurred with its recommendation
\cite{HHSOIG2026HospiceEligibility}. Because the audit focused on a
specifically defined high-risk population, its findings should not be
generalized to all hospice enrollees. They nevertheless demonstrate that
growth, public financing, quality measurement, and program integrity can
no longer be treated as separate management concerns.

Provider ownership and oversight, however, explain only part of the system
that produces hospice outcomes. Medicare hospice care is delivered
predominantly in patients' homes and residential settings
\cite{CMS2026HospiceGeneralInfo}. Agencies must deploy nurses, aides,
social workers, physicians, counselors, and other professionals across
geographically dispersed locations; respond to urgent needs; coordinate
services across organizations; and support family caregivers who often
provide substantial day-to-day care. The feasibility of these tasks may
differ sharply across counties. Rurality, social vulnerability,
population health burden, travel requirements, and local workforce and
health-resource capacity can shape both the needs confronting an agency
and the resources available to meet them
\cite{AdayAndersen1974AccessFramework,Andersen1995BehavioralModel,
Flanagan2011SVI,Hart2005RuralDefinitions}.

This creates a central performance-assessment dilemma. Unadjusted
comparisons may penalize providers serving communities with greater
structural disadvantage and fewer complementary resources. Context,
however, cannot be allowed to explain away organizational shortcomings
that may be remediable. The same raw score may reflect different
combinations of governance, operational capacity, local need, and resource
constraint. Evidence from other health-care settings similarly shows that
organizations serving more deprived populations may have greater
resources or higher measured efficiency while still achieving lower
effectiveness \cite{Merelie2025DeprivationDEA}. The relevant question is
therefore not simply which hospices have low scores. It is which provider
characteristics remain associated with caregiver experience after
observable operating conditions are considered, and where weak experience
coincides with severe community vulnerability.

Existing research does not yet provide a unified national answer. Hospice
ownership studies have identified important differences across ownership
forms but have not generally integrated a broad set of county-level
vulnerability, rurality, health-burden, and workforce conditions.
Geographic and equity research often focuses on access or utilization
without preserving the hospice agency as the unit of quality measurement
and managerial action. As a result, poor caregiver experience may be
attributed too readily either to provider organization or to community
disadvantage when the two should be assessed together.

This study addresses that gap by developing a context-aware
provider--county performance assessment framework for U.S. hospice quality
and family caregiver experience. We link national CMS hospice
characteristics, ownership and enrollment information, utilization data,
and provider-level CAHPS outcomes with county measures of rurality, social
vulnerability, community health burden, and workforce and health-resource
capacity. The linked provider base contains 6,852 hospice agencies.
Caregiver ratings are available for 3,178 providers, and the preferred
adjusted analysis includes 2,928 providers located in 1,078 counties. The
analysis combines geographic overlap assessment, blockwise adjusted
regression, six secondary CAHPS outcome models, and sensitivity analyses.
Because the design is observational and cross-sectional, the findings are
interpreted as adjusted associations for performance assessment rather
than causal effects.

The contribution is not another unadjusted ranking of hospice scores. It
is an empirical structure for separating, as far as national public data
permit, provider organization from the local conditions under which
home-based end-of-life care is delivered. By retaining the hospice agency
as the unit of accountability while incorporating county context, the
framework evaluates whether ownership and other provider characteristics
remain associated with family experience after adjustment and identifies
where poor experience overlaps with high vulnerability. This supports a
shift from one-size-fits-all comparison toward context-aware monitoring,
targeted quality improvement, and more defensible resource planning.

The remainder of the article is organized as follows.
Section~\ref{sec:literature} develops the conceptual foundation linking
provider organization, caregiver experience, and county operating
context. Section~\ref{sec:empirical} describes the data linkage, variable
construction, and empirical strategy. Section~\ref{sec:results} presents
the geographic, regression, secondary-outcome, and sensitivity results.
Section~\ref{sec:implications} discusses the managerial and policy
implications, and Section~\ref{sec:conclusion} concludes.

% ============================================================
\section{Literature Review}
\label{sec:literature}
% ============================================================

A context-aware assessment of hospice performance rests on four strands of
research: ownership and governance, county operating context,
caregiver-reported experience, and health-care performance assessment.
Taken together, these literatures show why the hospice agency should remain
the unit of accountability while its outcomes are interpreted in light of
the local conditions under which home-based care is delivered.

\subsection{Hospice ownership, governance, and caregiver experience}
\label{subsec:ownership_literature}

Ownership can shape hospice performance through governance, capital
allocation, staffing priorities, service intensity, admission and
discharge practices, organizational scale, and investment in quality
improvement. These channels matter because hospice agencies retain
considerable discretion over how interdisciplinary services are organized,
how personnel are deployed, and how patients and families are supported.
Ownership is therefore more than an administrative classification. It may
capture differences in decision authority, financial incentives,
accountability, and organizational priorities that affect the care process.

Recent studies underscore both the importance and the complexity of these
relationships. Aldridge et al. identified 124 U.S. hospice acquisitions by
47 private-equity firms between 2015 and 2022 and showed that complex
ownership arrangements are not adequately represented in standard public
classifications \cite{Aldridge2024HospicePEOpacity}. Using 2021--2022
CAHPS Hospice data, Soltoff et al. found that hospices owned by
private-equity firms or publicly traded companies performed less favorably
across caregiver-reported domains than nonprofit and other for-profit
agencies after adjustment for measured hospice and community
characteristics \cite{Soltoff2024JAMAHospiceOwnership}. Their analysis
also highlights a practical limitation of public reporting: CAHPS data
cover a large share of Medicare hospice users but only about half of active
hospice agencies.

These findings do not imply that every for-profit hospice performs poorly
or that ownership affects performance through a single mechanism.
For-profit entry may alter access, utilization, length of stay, spending,
and service organization in addition to caregiver experience. Gruber et
al. provide causal evidence that for-profit hospice entry increased
hospice use among some beneficiaries with Alzheimer's disease and related
dementias and reduced Medicare spending by substituting hospice for more
costly care \cite{Gruber2025DyingOrLying}. An ownership model may therefore
expand access or reduce expenditure while raising separate questions about
staffing, service mix, transparency, or family experience.

The empirical issue is not whether one ownership category should be
presumed better or worse. It is whether observed ownership differences
remain associated with caregiver experience after relevant provider
characteristics and county operating conditions are taken into account.
This distinction is especially important here because national CMS files
identify broad categories such as nonprofit, for-profit, government, and
other ownership, but do not consistently identify every private-equity or
publicly traded relationship. Research on private-equity ownership
therefore informs the governance rationale, whereas the primary analysis
evaluates the broader ownership categories observable in national public
data.

\subsection{County operating context and home-based hospice delivery}
\label{subsec:county_context_literature}

Hospice outcomes are produced within local service environments.
Access-to-care frameworks and neighborhood-context research show that
observed health-care outcomes reflect not only provider behavior but also
geography, population need, socioeconomic conditions, and resource
availability
\cite{AdayAndersen1974AccessFramework,Andersen1995BehavioralModel,
Flanagan2011SVI,KindBuckingham2018NeighborhoodAtlas,
Hart2005RuralDefinitions}. This perspective is particularly relevant to
hospice because care is delivered largely in patients' homes and
residential settings rather than within a single controlled facility.

County social vulnerability may increase both the complexity of care needs
and the difficulty of meeting them. Economic insecurity, transportation
barriers, unstable housing, limited social support, language and
communication barriers, and reduced access to complementary services can
increase the burden on family caregivers and hospice teams. Rurality may
add longer travel distances, dispersed demand, thinner labor markets, and
limited referral or specialty support. Community health burden may reflect
greater patient and caregiver needs, while local workforce and
health-resource measures may indicate the labor and service environment
from which hospices recruit personnel and coordinate care.

These contextual measures require disciplined interpretation. A
county-level workforce-shortage indicator does not measure staffing within
an individual hospice. A county health-burden index likewise does not
measure the clinical complexity of a hospice's own patients. Rather, these
variables describe the operating context in which provider performance is
observed.

This distinction guards against two common errors. The first is to assign
all observed variation to provider organization while ignoring the
conditions under which agencies operate. The second is to describe
disadvantaged places without preserving the provider-level outcomes on
which managers, regulators, and quality-improvement programs can act. A
provider--county framework addresses both concerns by retaining the hospice
agency as the unit of performance assessment while incorporating the local
conditions that may shape service delivery.

Recent evidence supports this approach. Merelie et al. found that
primary-care organizations serving more deprived populations could have
greater resources per capita and higher measured efficiency while
remaining less effective than organizations serving less deprived
populations \cite{Merelie2025DeprivationDEA}. Resource availability,
efficiency, and effectiveness therefore need not move together. For
hospice care, the implication is not that deprivation determines quality,
but that similar raw outcomes may arise from different combinations of
organizational capability and environmental constraint.

\subsection{Caregiver experience as an operational management outcome}
\label{subsec:caregiver_experience_literature}

Caregiver-reported outcomes are especially informative in hospice because
family members often observe the continuity and responsiveness of service
delivery across settings and over time. They experience whether calls are
answered, whether staff arrive when promised, whether symptoms are
addressed, whether instructions are understandable, and whether the
hospice prepares the family for changes in the patient's condition.

The CAHPS Hospice Survey captures dimensions of care that claims data and
process measures alone cannot fully represent. Its domains include
communication, timely help, respectful treatment, emotional and spiritual
support, symptom management, caregiver training, overall rating, and
willingness to recommend
\cite{Teno2004FamilyPerspectives,Kelley2015PalliativeCare,
Elliott2009CAHPSAdjustment,AHRQ2026CAHPSHospice}. These measures should not
be reduced to generic satisfaction scores. Communication and caregiver
training reflect the quality of information transfer. Timely help reflects
responsiveness and deployable capacity. Symptom support reflects
coordination between assessment and intervention. Overall rating captures
the family's integrated experience of the service system.

Caregiver-reported outcomes therefore provide a service-level view of
whether organizational resources are translated into accessible,
coordinated, and supportive care. Their multidomain structure also has
analytical value. An association found only for overall rating may reflect
a broad impression rather than a specific operational pathway. A pattern
that recurs across timely care, communication, symptom support, respect,
caregiver training, and willingness to recommend offers stronger evidence
that a provider or contextual characteristic is related to several
components of service delivery. This logic supports the use of overall
hospice rating as the primary outcome and six additional CAHPS domains as
secondary outcomes.

\subsection{Context-aware health-care performance assessment}
\label{subsec:performance_literature}

Health-care management science views quality as an outcome of the system
that produces it. Staffing research links workload and nursing capacity to
mortality, burnout, turnover, and cost
\cite{Aiken2002NurseStaffing,Kane2007NurseStaffingOutcomes,
Cho2023PatientNurseRatios}. Queueing-based work shows that the value of
additional staffing depends on patient mix, workload, and the delay before
care arrives \cite{Sulz2024PatientBasedNurseStaffing}. Capacity-planning
and patient-assignment studies likewise show that geography, demand
heterogeneity, workforce constraints, equity, uncertainty, and stakeholder
needs should be considered jointly rather than through a single uniform
rule
\cite{Khalili2025LTCCapacityPlanning,Schafer2023PatientBedAssignment}.

These studies are not invoked to characterize the present analysis as a
prescriptive optimization model. Their relevance lies in a more basic
management principle: performance should be interpreted in relation to the
demand, capacity, and operating conditions that produce it.

The benchmarking literature reaches a similar conclusion. Data
Envelopment Analysis and related methods evaluate decision-making units
relative to their inputs, outputs, peers, and feasible operating
conditions
\cite{Charnes1978DEA,Banker1984DEA,Andersen1993SuperEfficiency,
Thanassoulis2001DEA,Dyson2001DEAPitfalls,
Podinovski2004WeightRestrictions}. Group-performance methods further
separate internal managerial performance from differences associated with
programs, policies, or environmental conditions
\cite{Camanho2006Malmquist}. The present study follows this
context-sensitive benchmarking logic but uses regression-based
provider--county assessment rather than an efficiency frontier. This
choice fits both the outcome structure and the empirical objective: CAHPS
measures are provider-level percentages, and the central task is to
estimate adjusted associations among caregiver experience, provider
characteristics, and county operating context.

The remaining gap is both conceptual and managerial. Ownership studies
have not generally integrated broad measures of rurality, social
vulnerability, community health burden, and local workforce capacity into
a national provider-level caregiver-experience framework. Geographic and
equity studies often emphasize access or utilization without retaining the
hospice agency as the unit of quality measurement and managerial action.
Operations and benchmarking research explains why context matters, but
hospice-specific applications linking governance, caregiver experience,
and county operating conditions remain limited.

The provider--county design developed in this study addresses that gap. It
retains the hospice agency as the level at which quality is reported and
accountability is exercised, while incorporating the county as the context
in which home-based care is organized. The design supports two
complementary interpretations: whether provider characteristics remain
associated with caregiver experience after observable county context is
considered, and where poor caregiver experience coincides with vulnerable
operating conditions. Section~\ref{sec:empirical} operationalizes this
framework.

% ============================================================
\section{Empirical Analysis}
\label{sec:empirical}
% ============================================================

This section translates the provider--county framework into an empirical design. The hospice agency remains the unit of performance assessment because caregiver-experience scores and managerial accountability are reported at the provider level. County measures are added to characterize the operating conditions under which home-based care is organized, including rurality, social vulnerability, community health burden, and local workforce and health-resource context. The analysis then combines linkage validation, geographic description, blockwise adjusted regression, secondary-outcome models, and sensitivity analyses. This structure preserves a clear distinction between provider organization and operating context while producing results that can be used for monitoring, peer comparison, and quality-improvement prioritization.

\subsection{Data sources and study population}

\begin{table}[!htbp]
\centering
\caption{Public data sources and analytical roles}
\label{tab:data_sources}
\scriptsize
\setlength{\tabcolsep}{3pt}
\renewcommand{\arraystretch}{1.08}
\begin{tabular}{L{0.30\textwidth}L{0.15\textwidth}C{0.10\textwidth}L{0.35\textwidth}}
\toprule
Data source & Unit & Year & Analytical role \\
\midrule
CMS Hospice General Information \cite{CMS2026HospiceGeneralInfo} & Provider & May 2026 & Provider base, address, ownership type, and certification date \\
CMS Hospice Provider Data \cite{CMS2026HospiceProviderData} & \makecell[l]{Provider-\\measure} & May 2026 & Hospice quality measures, provider scale, case mix, and care setting \\
Provider CAHPS Hospice Survey Data \cite{CMS2026CAHPSHospiceProvider,AHRQ2026CAHPSHospice} & \makecell[l]{Provider-\\measure} & May 2026 & Family caregiver experience outcomes \\
CMS Hospice ZIP file \cite{CMS2026HospiceZip} & \makecell[l]{Provider-\\ZIP} & May 2026 & Service ZIP areas and access context \\
Hospice All Owners \cite{CMS2026HospiceAllOwners} & \makecell[l]{Provider/\\owner} & 2026 Q2 & Ownership structure and owner-level characteristics \\
Hospice Enrollments \cite{CMS2026HospiceEnrollments} & \makecell[l]{Provider/\\enrollment} & 2026 Q2 & Enrollment identity, National Provider Identifier (NPI), and enrollment validation \\
Medicare Post-Acute Care (PAC) Hospice Utilization \cite{CMS2023PACHospiceUtilization} & \makecell[l]{Provider/\\geography} & 2023 & Utilization, payment, and service-intensity context \\
HRSA Area Health Resources File \cite{HRSA2025AHRF} & County & 2024--2025 & Workforce shortage and health-resource capacity proxies \\
USDA Rural--Urban Continuum Codes \cite{USDA2023RUCC} & County & 2023 & Rurality and metropolitan status \\
CDC/ATSDR Social Vulnerability Index \cite{CDC2022SVI,Flanagan2011SVI} & County & 2022 & Social vulnerability and deprivation context \\
CDC PLACES \cite{CDC2025PLACESCounty,CDC2025PLACESDefinitions} & County & 2025 & Community health burden and health-risk prevalence \\
\bottomrule
\end{tabular}
\vspace{0.5em}
\begin{minipage}{0.95\textwidth}
\footnotesize\emph{Note.} All sources are public administrative, survey, or county-level contextual data sources. Provider-level files are linked by CMS Certification Number (CCN); county-level files are linked by county Federal Information Processing Standards (FIPS) code.
\end{minipage}
\end{table}

Table~\ref{tab:data_sources} summarizes the public data sources used in the study and their analytical roles. The provider base is the CMS Hospice General Information file released in May 2026. Provider-level quality measures come from the CMS Hospice Provider Data file, and family caregiver experience outcomes come from the provider-level CAHPS Hospice Survey file. Additional provider and organizational information is obtained from hospice ownership and enrollment data, while Medicare post-acute care utilization data provide service-utilization, beneficiary, and payment context. County-level contextual data include the U.S. Department of Agriculture (USDA) Rural--Urban Continuum Codes (RUCC) 2023; the Centers for Disease Control and Prevention/Agency for Toxic Substances and Disease Registry (CDC/ATSDR) Social Vulnerability Index (SVI) 2022; CDC Population Level Analysis and Community Estimates (PLACES) 2025; and the Health Resources and Services Administration (HRSA) Area Health Resources Files (AHRF) 2024--2025. The resulting design links the unit of accountability, the hospice provider, with the local operating conditions under which hospice care is delivered.

\begin{landscape}
\begin{figure}[p]
\centering
\makebox[\linewidth][c]{\includegraphics[width=1.0\linewidth]{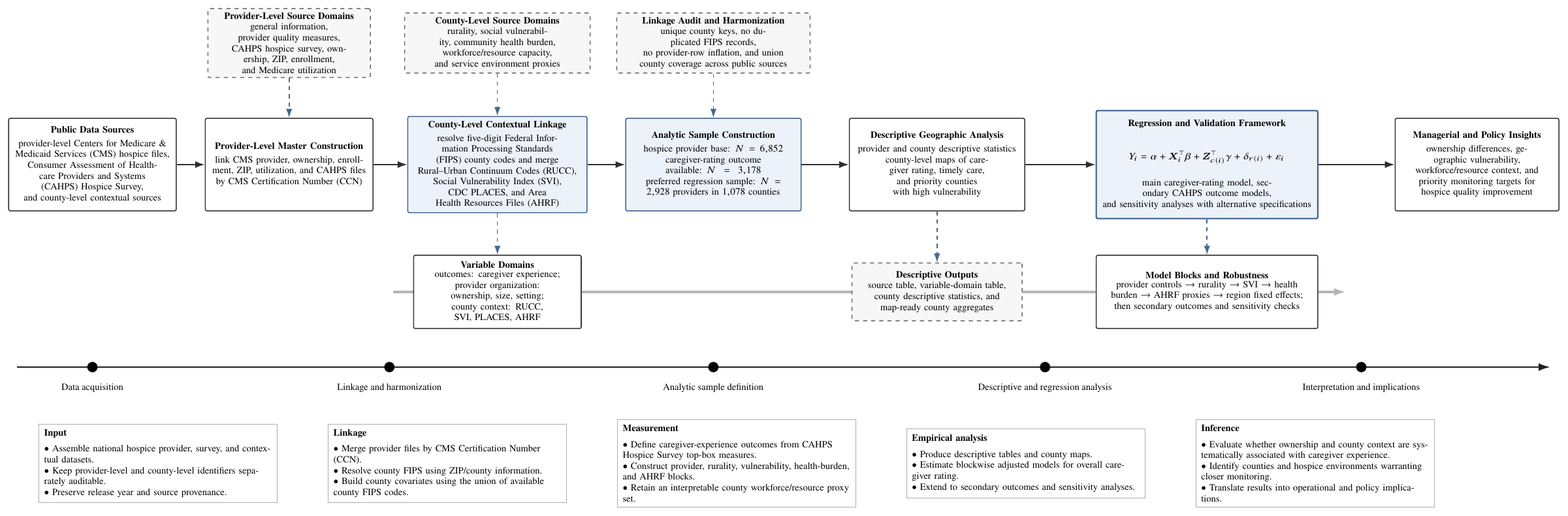}}
\caption{Study design, data linkage, and context-aware hospice performance assessment framework. Provider-level Centers for Medicare \& Medicaid Services (CMS) hospice files and Consumer Assessment of Healthcare Providers and Systems (CAHPS) Hospice Survey outcomes are linked by CMS Certification Number (CCN). County-level contextual measures are harmonized using Federal Information Processing Standards (FIPS) county codes. The workflow proceeds from data assembly and linkage validation to geographic analysis, adjusted regression modeling, secondary-outcome validation, sensitivity analyses, and operational interpretation.}
\label{fig:study_workflow}
\end{figure}
\end{landscape}

Figure~\ref{fig:study_workflow} summarizes the study design, data-linkage sequence, and empirical workflow. The figure serves as a reader-facing guide to how provider-level CMS and CAHPS Hospice files are linked to county-level contextual sources, audited for linkage validity, analyzed through geographic and regression models, and translated into management and policy insights.

The CMS Hospice General Information file contains 6,852 hospice providers. Provider-level CMS quality measures are available for 6,312 providers after pivoting measure-level data to a provider-level wide format. Provider-level CAHPS Hospice scores are available for 3,178 providers. County FIPS codes are resolved for 6,686 providers after District of Columbia correction. Among providers with resolved county FIPS codes, match rates to SVI, PLACES, and AHRF county data exceed 99.9\%.

County-level covariates were harmonized using five-digit county FIPS codes. Because national public data sources do not define identical county universes, the county-level covariate table was constructed as the union of available county FIPS codes across RUCC, SVI, PLACES, and AHRF. This process produced 3,242 unique county-level records. The nine-record difference between the initial county base and the final county-covariate universe reflects source-specific county coverage rather than duplicate linkage. Linkage validation confirmed that all county FIPS identifiers were unique within the final county covariate table, that the AHRF source file contained unique county FIPS identifiers, and that provider-level linkage preserved all 6,852 hospice provider records.

\begin{table}[!htbp]
\centering
\caption{Analytic sample construction and linkage success}
\label{tab:sample_construction}
\begin{tabular}{lr}
\toprule
Sample construction step & Number of providers \\
\midrule
CMS Hospice General Information provider base & 6,852 \\
With CMS Hospice quality measures & 6,312 \\
With provider-level CAHPS Hospice scores & 3,178 \\
With resolved county FIPS after DC correction & 6,686 \\
With SVI match among providers with resolved FIPS & 6,683 \\
With PLACES 2025 match among providers with resolved FIPS & 6,683 \\
With AHRF 2024--2025 match among providers with resolved FIPS & 6,685 \\
Final regression sample for primary adjusted model & 2,928 \\
\bottomrule
\end{tabular}
\end{table}

\subsection{Measures and variable construction}

\subsubsection{Family caregiver experience outcomes}

Let \(i\) index hospice providers, \(c(i)\) denote the county in which provider \(i\) is located, and \(r(i)\) denote the CMS region or state associated with provider \(i\). For each CAHPS domain \(k\), let \(Y_i^{(k)}\) denote the provider-level caregiver experience score. The primary outcome is the percentage of caregivers rating the hospice agency a 9 or 10, denoted by \(Y_i^{(\mathrm{rating})}\). Secondary outcomes include willingness to recommend the hospice agency, timely care, team communication, symptom support, respect, and emotional/spiritual support.

\subsubsection{Provider-level characteristics and ownership}

Provider-level covariates \(X_i\) include ownership type, certification date, average daily census, care setting, service type, payer mix, and patient mix variables derived from CMS Hospice Provider Data and related provider files. The initial analysis uses the ownership type reported in CMS Hospice General Information. More detailed owner-level variables from Hospice All Owners are summarized separately before inclusion to avoid row expansion from many-to-one owner records. Because recent hospice ownership studies distinguish private-equity/publicly traded company ownership from other for-profit ownership \cite{Soltoff2024JAMAHospiceOwnership,Aldridge2024HospicePEOpacity}, our main public-data specification treats CMS ownership type as the baseline ownership variable and considers owner-file-derived ownership complexity measures as secondary analyses when link quality is confirmed.

\subsubsection{County-level contextual measures}

County-level covariates \(Z_{c(i)}\) include rurality from RUCC, social vulnerability from SVI, community health burden from PLACES, and local workforce and health-resource capacity from AHRF. SVI measures include the overall vulnerability percentile and theme-specific percentiles. PLACES measures include adjusted prevalence of chronic conditions, disability, poor physical health, poor mental health, lack of health insurance, and related community health-burden indicators. AHRF variables are used as proxies for local workforce and health-resource capacity, not as direct provider staffing measures. To avoid overparameterizing the county-level resource environment, the main model uses a parsimonious AHRF block covering advanced-practice nursing availability, primary-care shortage context, mental-health shortage context, and local health-system infrastructure capacity. Exact source variable names are preserved in the audit files and analysis code; the manuscript reports descriptive labels to keep the main text readable. Because these measures are county-level variables, they are interpreted as the broader local workforce and health-resource environment in which hospice agencies operate, rather than as hospice-specific staffing levels.

\begin{table}[!htbp]
\centering
\caption{Variable domains used in the analysis}
\label{tab:variables}
\scriptsize
\setlength{\tabcolsep}{3pt}
\renewcommand{\arraystretch}{1.08}
\begin{tabular}{L{0.17\textwidth}L{0.35\textwidth}L{0.15\textwidth}L{0.19\textwidth}}
\toprule
Domain & Example variables & Source & Role \\
\midrule
Caregiver experience & Overall caregiver rating, willingness to recommend, timely care, team communication, symptom management, respect, and emotional/spiritual support & CAHPS Hospice & Outcomes \\
Provider organization & Ownership type, certification date, provider scale, and care setting & CMS Hospice & Predictors/controls \\
Hospice quality and case mix & Average daily census, payer mix, disease mix, and care location & CMS Hospice Provider Data & Predictors/controls \\
Utilization and payment & Service utilization, payment, and beneficiary characteristics & Medicare PAC & Controls/sensitivity \\
Rurality & RUCC code and metro/nonmetro categories & USDA RUCC & County context \\
Social vulnerability & Overall SVI percentile and theme percentiles & CDC/ATSDR SVI & County context \\
Community health burden & Chronic disease, disability, poor health, and insurance access & CDC PLACES & County context \\
Workforce/resource capacity & Advanced-practice nursing availability, primary-care shortage, mental-health shortage, and local health-system infrastructure capacity & HRSA AHRF & County context \\
\bottomrule
\end{tabular}
\end{table}

\subsection{Formal provider--county performance model}

The central empirical model is a provider--county performance model:
\begin{equation}
Y_i^{(k)} = \alpha^{(k)} + X_i^\top \beta^{(k)} + Z_{c(i)}^\top \gamma^{(k)} + \delta_{r(i)}^{(k)} + \varepsilon_i^{(k)},
\label{eq:provider_county_model}
\end{equation}
where \(Y_i^{(k)}\) is the caregiver experience outcome for domain \(k\), \(X_i\) is a vector of provider-level characteristics, \(Z_{c(i)}\) is a vector of county-level contextual variables, \(\delta_{r(i)}^{(k)}\) captures CMS-region or state fixed effects, and \(\varepsilon_i^{(k)}\) is the residual provider-level variation.

This formulation separates three components of hospice performance:
\begin{equation}
Y_i^{(k)}
=
\underbrace{X_i^\top \beta^{(k)}}_{\text{provider organization}}
+
\underbrace{Z_{c(i)}^\top \gamma^{(k)}}_{\text{community context}}
+
\underbrace{\delta_{r(i)}^{(k)}}_{\text{regional environment}}
+
\varepsilon_i^{(k)}.
\label{eq:decomposition}
\end{equation}
This decomposition is not interpreted causally; rather, it provides a structured way to quantify how provider organization and county operating context are jointly associated with observed hospice caregiver experience. This distinction is important because the managerial unit is the provider, whereas many access, workforce, and vulnerability constraints are observed at the county level.

\subsection{Blockwise explanatory contribution}

To evaluate the incremental explanatory role of each contextual domain, we estimate a sequence of nested, interpretable models that mirrors the specification structure used in the main results table:
\begin{align}
\mathcal{M}_1 &: Y_i^{(k)} \sim X_i, \\
\mathcal{M}_2 &: Y_i^{(k)} \sim X_i + \mathrm{RUCC}_{c(i)}, \\
\mathcal{M}_3 &: Y_i^{(k)} \sim X_i + \mathrm{RUCC}_{c(i)} + \mathrm{SVI}_{c(i)}, \\
\mathcal{M}_4 &: Y_i^{(k)} \sim X_i + \mathrm{RUCC}_{c(i)} + \mathrm{SVI}_{c(i)} + \mathrm{PLACES}_{c(i)}, \\
\mathcal{M}_5 &: Y_i^{(k)} \sim X_i + \mathrm{RUCC}_{c(i)} + \mathrm{SVI}_{c(i)} + \mathrm{PLACES}_{c(i)} + \mathrm{AHRF}_{c(i)}, \\
\mathcal{M}_6 &: Y_i^{(k)} \sim X_i + \mathrm{RUCC}_{c(i)} + \mathrm{SVI}_{c(i)} + \mathrm{PLACES}_{c(i)} + \mathrm{AHRF}_{c(i)} + \delta_{r(i)}.
\end{align}
For block \(b\), the incremental explanatory contribution is summarized by the change in adjusted explanatory fit:
\begin{equation}
\Delta \bar{R}_b^2 = \bar{R}^2(\mathcal{M}_b) - \bar{R}^2(\mathcal{M}_{b-1}), \quad b=2,\ldots,6.
\label{eq:incremental_r2}
\end{equation}
This blockwise structure is used for interpretation rather than model selection alone. It shows whether rurality, social vulnerability, community health burden, workforce/resource capacity, and regional environment add meaningful explanatory information after provider organization has been accounted for. In this sense, the sequence provides a regression-based analogue to the group-performance logic in benchmarking studies: rather than estimating a frontier directly, we quantify how successive operating-context domains change the interpretation of provider-level caregiver experience.

\subsection{Context-adjusted performance and benchmarking}

For each provider \(i\), the fitted value from Eq.~\eqref{eq:provider_county_model} is
\begin{equation}
\widehat{Y}_i^{(k)}
=
\widehat{\alpha}^{(k)}
+
X_i^\top \widehat{\beta}^{(k)}
+
Z_{c(i)}^\top \widehat{\gamma}^{(k)}
+
\widehat{\delta}_{r(i)}^{(k)}.
\end{equation}
The context-adjusted residual performance measure is defined as
\begin{equation}
R_i^{(k)} = Y_i^{(k)} - \widehat{Y}_i^{(k)}.
\label{eq:residual_performance}
\end{equation}
Providers with positive residuals perform above model-expected levels given their observed provider and county context; providers with negative residuals perform below model-expected levels. This residual is used for context-aware benchmarking and to identify providers that may serve as learning peers under comparable operating conditions. For example, a provider can be compared with agencies facing similar vulnerability and resource environments rather than with an unadjusted national average. The residual-benchmarking approach is deliberately conservative: it does not claim technical efficiency from CAHPS outcomes alone, but it preserves the benchmarking intent of DEA and super-efficiency approaches by asking which providers appear to perform better or worse than expected under observed operating conditions \cite{Charnes1978DEA,Banker1984DEA,Andersen1993SuperEfficiency}.

\subsection{County-level aggregation and priority counties}

To visualize geographic variation, provider outcomes are aggregated to the county level:
\begin{equation}
\bar{Y}_c^{(k)} = \frac{1}{n_c}\sum_{i:c(i)=c}Y_i^{(k)},
\label{eq:county_mean}
\end{equation}
where \(n_c\) is the number of hospice providers with non-missing outcome \(k\) in county \(c\).

We define priority counties as those with low caregiver experience and high social vulnerability:
\begin{equation}
\mathcal{P}^{(k)}
=
\left\{
c :
\bar{Y}_c^{(k)} \leq Q_{25}\left(\bar{Y}^{(k)}\right),
\;
\mathrm{SVI}_c \geq Q_{75}\left(\mathrm{SVI}\right)
\right\}.
\label{eq:priority_set}
\end{equation}
The set \(\mathcal{P}^{(k)}\) identifies counties where caregiver experience is in the bottom quartile while social vulnerability is in the top quartile. These counties are interpreted as priority areas for management attention, quality-improvement support, and policy monitoring, not as deterministic labels of provider quality.

A complementary four-quadrant classification is also used:
\begin{equation}
G_c^{(k)} =
\begin{cases}
\mathrm{HighExperience/LowVulnerability}, & \bar{Y}_c^{(k)} > Q_{50}(\bar{Y}^{(k)}),\ \mathrm{SVI}_c \le Q_{50}(\mathrm{SVI}),\\
\mathrm{HighExperience/HighVulnerability}, & \bar{Y}_c^{(k)} > Q_{50}(\bar{Y}^{(k)}),\ \mathrm{SVI}_c > Q_{50}(\mathrm{SVI}),\\
\mathrm{LowExperience/LowVulnerability}, & \bar{Y}_c^{(k)} \le Q_{50}(\bar{Y}^{(k)}),\ \mathrm{SVI}_c \le Q_{50}(\mathrm{SVI}),\\
\mathrm{LowExperience/HighVulnerability}, & \bar{Y}_c^{(k)} \le Q_{50}(\bar{Y}^{(k)}),\ \mathrm{SVI}_c > Q_{50}(\mathrm{SVI}).
\end{cases}
\label{eq:quadrant}
\end{equation}
This classification is designed for bivariate mapping and for identifying resilient counties that achieve above-median caregiver experience despite high social vulnerability.

\subsection{Empirical models and inference}

The primary regression models are estimated using ordinary least squares for continuous percentage outcomes. Models include CMS-region or state fixed effects depending on sample size and collinearity diagnostics. Standard errors are clustered at the county level when feasible; alternative robust standard errors are evaluated in sensitivity analyses. This specification is chosen for transparency and interpretability rather than for causal identification. Because the study is cross-sectional and based on public data releases, estimates are interpreted as adjusted associations rather than causal effects.

\subsection{Robustness and sensitivity analyses}

Robustness analyses evaluate whether the main caregiver-rating findings are sensitive to fixed-effect choice, contextual block inclusion, sparse-county observations, outcome definition, and public-reporting sample restrictions. Specifically, we estimate: (i) the preferred caregiver-rating model; (ii) a state fixed-effects specification in place of CMS region fixed effects; (iii) a model excluding the PLACES health-burden index; (iv) a model excluding the AHRF workforce/resource block; (v) a reduced AHRF specification using only shortage-context measures; (vi) a sample restricted to counties with at least two CAHPS-rating providers; (vii) a sample restricted to providers with nonmissing summary star rating; and (viii) an alternative primary-outcome model using timely care. These specifications are designed to test stability rather than to support causal interpretation.

% ============================================================

% ============================================================
\section{Results}
\label{sec:results}
% ============================================================

The results are presented in the same order as the assessment framework. We first verify the provider--county linkage and describe the analytic sample. We then show geographic variation in caregiver experience and priority counties. Finally, we estimate the main blockwise caregiver-rating models and test whether the central associations are stable across secondary CAHPS outcomes and sensitivity specifications. This ordering is intended to make the empirical evidence easy to audit: the maps show where differences appear, the blockwise models show what changes after adding each operating-context domain, and the validation analyses show whether the core findings persist beyond a single outcome or specification.

\subsection{Sample construction and data linkage}

Table~\ref{tab:sample_construction} reports the construction of the analytic samples. The provider base includes 6,852 hospice providers. Quality measures are available for 6,312 providers and provider-level CAHPS Hospice scores are available for 3,178 providers. After county name matching and District of Columbia correction, county FIPS codes are resolved for 6,686 providers. Among providers with resolved FIPS codes, county-level match rates are 99.96\% for SVI, 99.96\% for PLACES 2025, and 99.99\% for AHRF 2024--2025. This audit trail is important because the empirical results depend on preserving one provider row per CCN and avoiding row inflation during provider--county linkage. In management terms, the analysis keeps the hospice agency as the evaluated unit while adding county context only as an operating environment.

\subsection{Provider and county characteristics}

Table~\ref{tab:descriptive_statistics} summarizes the CAHPS analytic sample used for the caregiver-experience analysis. The sample contains 3,178 providers with nonmissing overall caregiver rating. County-level contextual variables are reported for providers whose county FIPS codes could be linked to SVI, PLACES, AHRF, and RUCC measures. This table should be read together with Figures~\ref{fig:rating_map}--\ref{fig:priority_map}: the descriptive statistics show the provider and county distributions, while the maps show how caregiver-experience and vulnerability patterns are distributed geographically. This pairing helps prevent overreading any single map as an adjusted performance ranking.

\begin{table}[tbp]
\centering
\caption{Descriptive statistics for the CAHPS analytic sample}
\label{tab:descriptive_statistics}
\begingroup
\fontsize{6.1}{6.9}\selectfont
\setlength{\tabcolsep}{1.5pt}
\renewcommand{\arraystretch}{0.98}
\begin{tabular}{p{0.36\textwidth}rrrrrr}
\toprule
Variable & Nonmissing & Missing (\%) & Mean & SD & Median & Range \\
\midrule
Family caregiver rating, top-box (\%) & 3178 & 0.00 & 81.882 & 6.623 & 83.000 & 51.000--99.000 \\
Timely care, top-box (\%) & 3178 & 0.00 & 77.721 & 7.271 & 78.000 & 44.000--98.000 \\
SVI overall percentile & 3113 & 2.05 & 0.579 & 0.268 & 0.622 & 0.005--0.998 \\
Average daily census & 3178 & 0.00 & 113.009 & 271.088 & 69.000 & 0.000--13421.000 \\
Care provided in home (\%) & 3178 & 0.00 & 55.027 & 22.725 & 56.000 & 0.000--100.000 \\
Care provided in assisted living (\%) & 3163 & 0.47 & 22.178 & 19.729 & 17.000 & 0.000--100.000 \\
COPD adjusted prevalence & 2940 & 7.49 & 6.025 & 1.366 & 5.800 & 3.200--11.000 \\
Coronary heart disease adjusted prevalence & 2940 & 7.49 & 5.478 & 0.711 & 5.500 & 3.800--8.200 \\
Diabetes adjusted prevalence & 2940 & 7.49 & 10.839 & 1.931 & 10.700 & 6.000--18.100 \\
Depression adjusted prevalence & 2940 & 7.49 & 23.138 & 3.454 & 23.000 & 13.000--36.200 \\
Disability adjusted prevalence & 2940 & 7.49 & 30.271 & 4.680 & 29.900 & 18.900--46.300 \\
Fair/poor general health adjusted prevalence & 2940 & 7.49 & 19.465 & 3.827 & 19.200 & 10.400--34.100 \\
AHRF advanced practice registered nurse (APRN) availability & 3113 & 2.05 & 1261.786 & 1894.307 & 495.000 & 0.000--9143.000 \\
AHRF primary-care shortage context & 3113 & 2.05 & 1.832 & 0.519 & 2.000 & 0.000--2.000 \\
AHRF mental-health shortage context & 3113 & 2.05 & 1.680 & 0.533 & 2.000 & 0.000--2.000 \\
AHRF local health-system infrastructure capacity & 3113 & 2.05 & 4.665 & 7.388 & 2.000 & 0.000--44.000 \\
RUCC 2023 code & 3113 & 2.05 & 2.445 & 1.929 & 2.000 & 1.000--9.000 \\
Nonmetro indicator & 3113 & 2.05 & 0.195 & 0.397 & 0.000 & 0.000--1.000 \\
\bottomrule
\end{tabular}
\endgroup

\vspace{0.35em}
\begin{minipage}{0.88\textwidth}
\scriptsize\emph{Note.} Descriptive statistics are calculated for providers with nonmissing CAHPS overall caregiver rating. AHRF variables are county-level workforce and health-resource environment proxies and should not be interpreted as hospice-specific staffing levels.
\end{minipage}
\end{table}

\subsection{Geographic variation in hospice caregiver experience}
\label{subsec:geographic_results}

\begin{figure}[!tbp]
\centering
\includegraphics[width=0.82\textwidth]{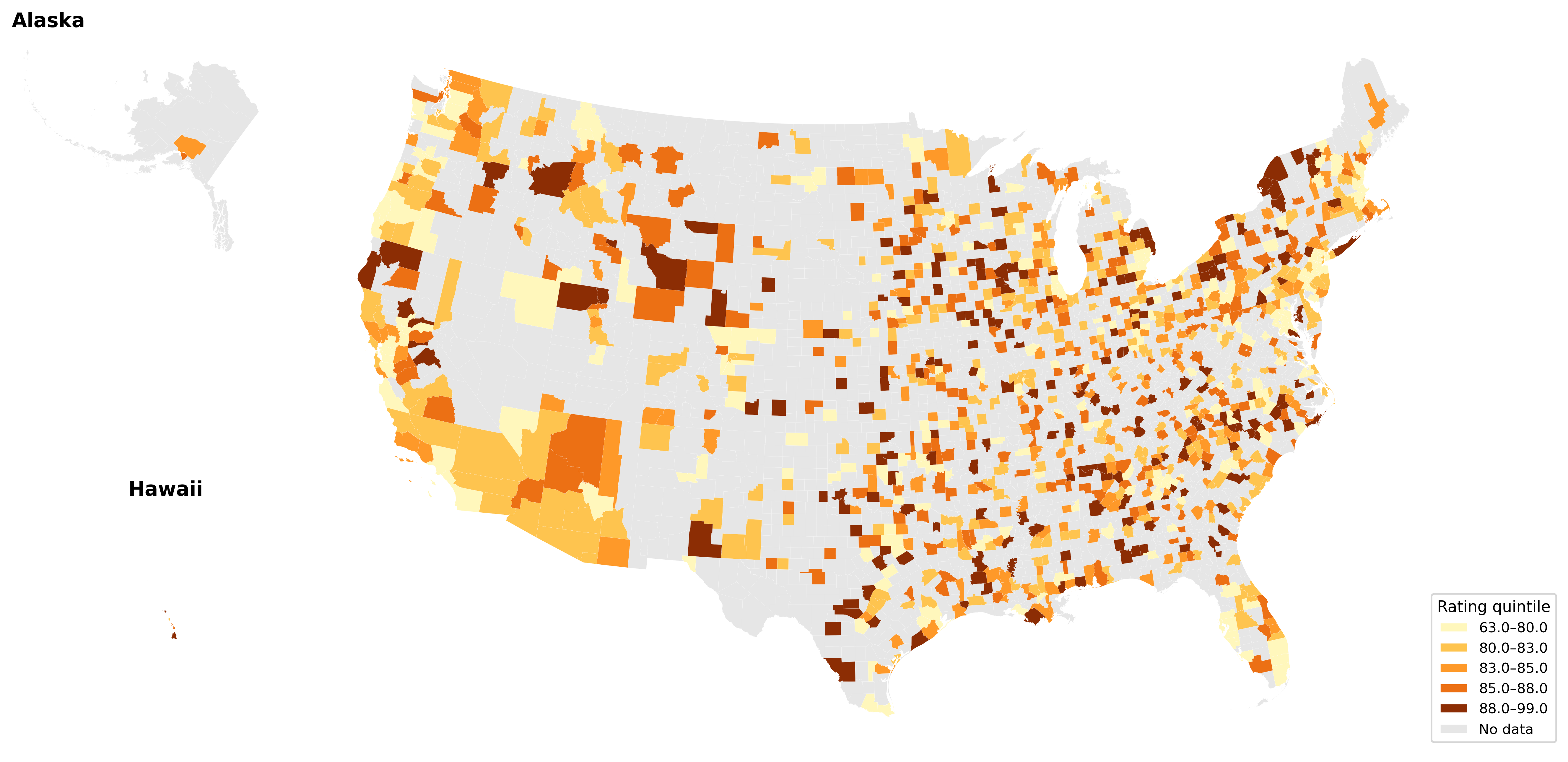}
\caption{County-level family caregiver rating. County means are computed from provider-level CAHPS Hospice rating top-box scores. Alaska and Hawaii are displayed outside the contiguous United States for geographic completeness.}
\label{fig:rating_map}
\end{figure}

\begin{figure}[!tbp]
\centering
\includegraphics[width=0.82\textwidth]{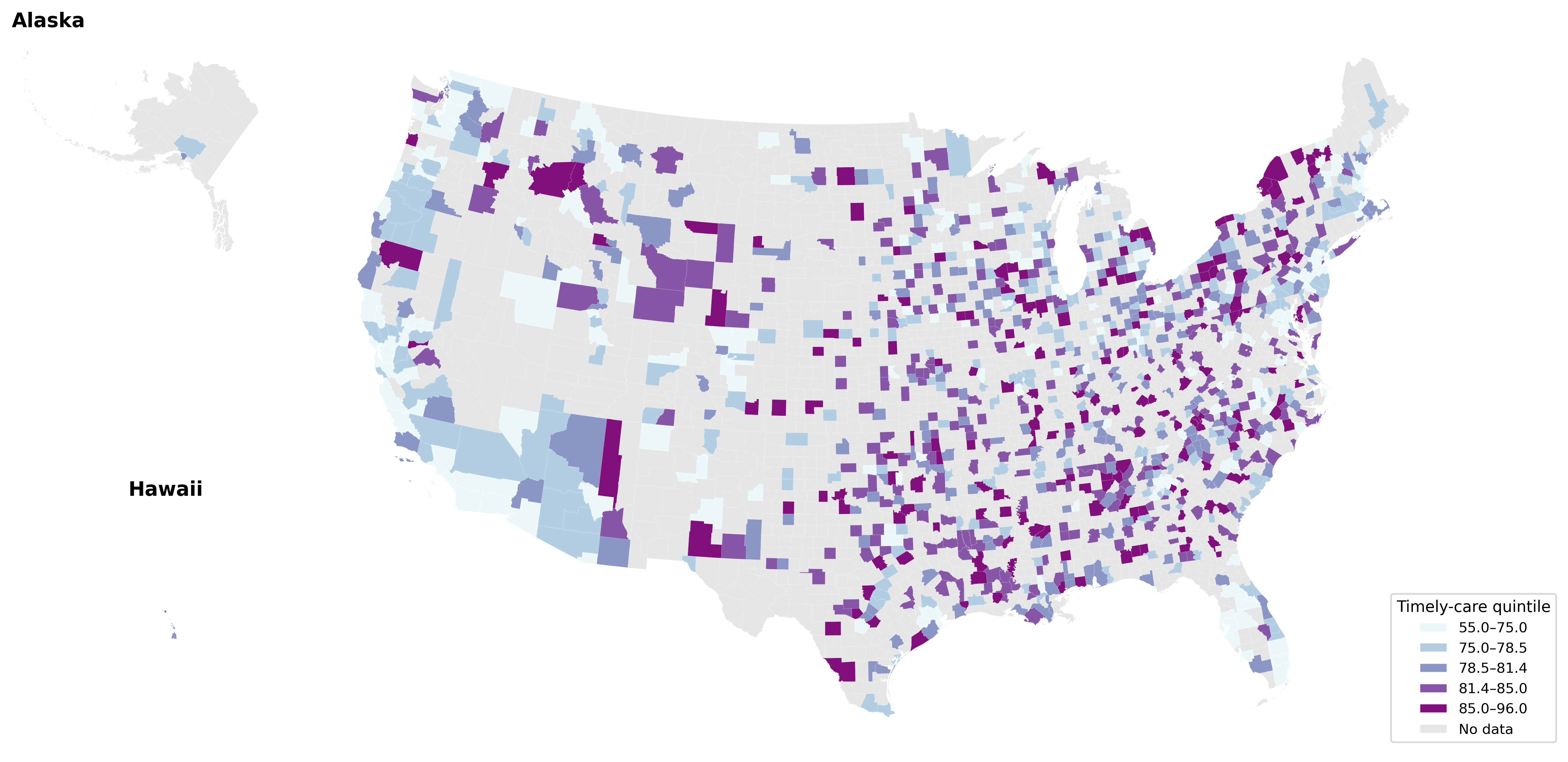}
\caption{County-level timely care. County means are computed from provider-level CAHPS Hospice timely-care top-box scores. Alaska and Hawaii are displayed outside the contiguous United States for geographic completeness.}
\label{fig:timely_map}
\end{figure}

\begin{figure}[!tbp]
\centering
\includegraphics[width=0.82\textwidth]{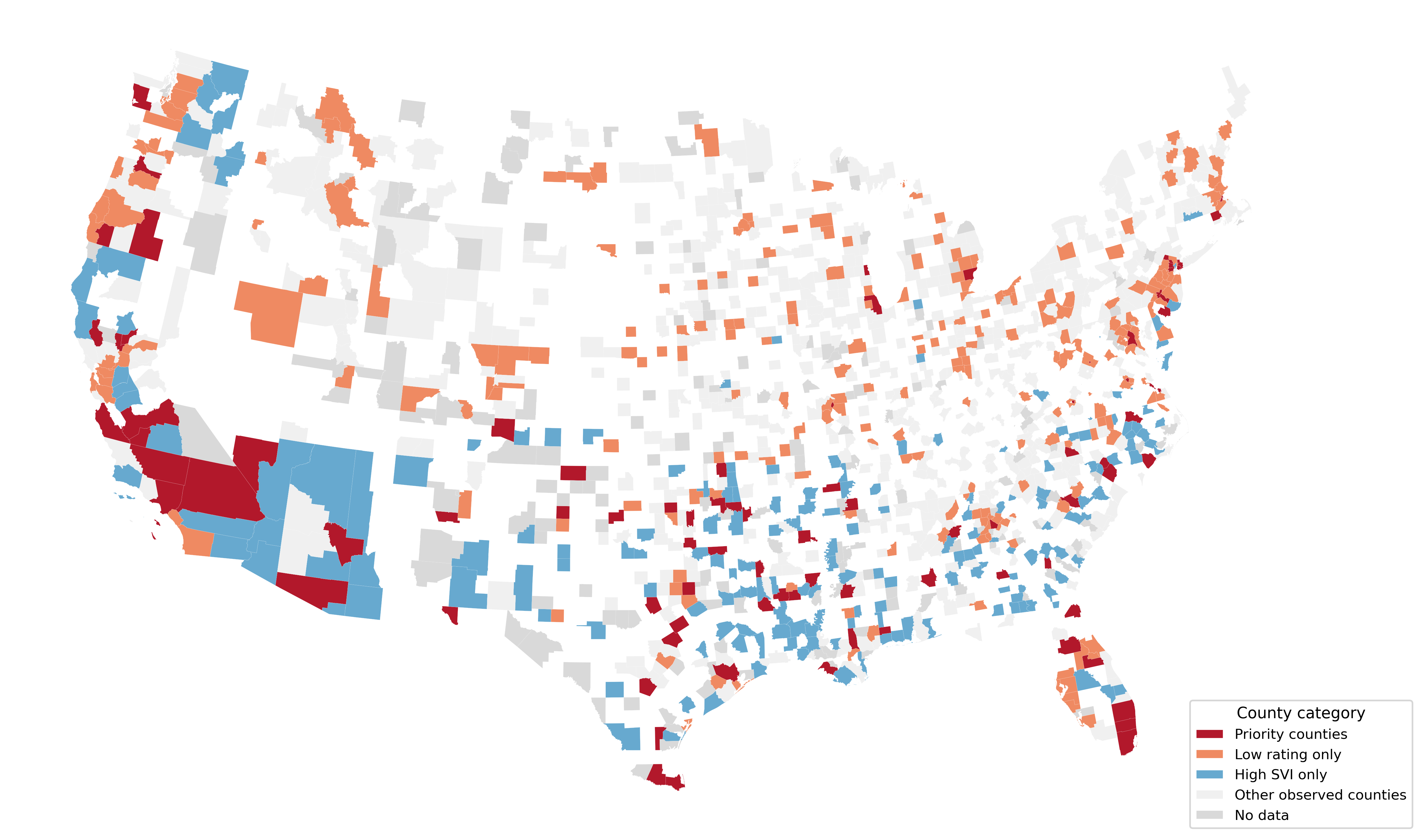}
\caption{Priority counties with low caregiver rating and high social vulnerability. Priority counties are defined as counties in the bottom quartile of county mean caregiver rating and the top quartile of county social vulnerability. Single-condition counties are shown for context. Alaska and Hawaii are omitted because their observed priority pattern is limited in this application and omission improves visual emphasis on the contiguous United States.}
\label{fig:priority_map}
\end{figure}

Figures~\ref{fig:rating_map}--\ref{fig:priority_map} provide three complementary views of the geographic structure of hospice caregiver experience. Constructing these maps required provider-level CAHPS Hospice outcomes to be linked to verified county Federal Information Processing Standards codes, aggregated without duplicating provider records, and then aligned with county social-vulnerability measures. The maps therefore do more than display raw survey scores. They translate a national provider-level reporting system into a geographic monitoring framework while preserving the hospice agency as the underlying unit from which county summaries are formed.

Figure~\ref{fig:rating_map} displays county means of the provider-level CAHPS Hospice overall caregiver-rating top-box score. This outcome captures the family's integrated assessment of the hospice experience and is the primary outcome used in the regression analysis. The map makes visible the spatial heterogeneity that is obscured by a national mean: counties represented in the public reporting data do not form a uniform performance landscape. At the same time, the figure is intentionally descriptive. County means may reflect differences in the mix of reporting providers, ownership, agency size, care setting, community need, and local resource conditions. The map therefore identifies where higher and lower observed caregiver ratings occur; it does not by itself determine why they occur or rank risk-adjusted provider performance.

Figure~\ref{fig:timely_map} maps the county mean of the CAHPS Hospice timely-care top-box score. Timely care is examined separately because it is more directly connected to operational responsiveness: whether help is available when families need it, whether calls and urgent needs are addressed, and whether geographically dispersed personnel can be deployed effectively. Comparing Figures~\ref{fig:rating_map} and~\ref{fig:timely_map} distinguishes a broad global assessment from a more specific service-delivery outcome. Areas with similar patterns across the two maps warrant attention because the observed signal is not confined to a single global rating item; differences between the maps are also informative because they may indicate that overall experience and operational responsiveness do not always move together.

Figure~\ref{fig:priority_map} combines performance and equity information that are otherwise shown separately. Counties are classified using two transparent national thresholds: the bottom quartile of county mean caregiver rating and the top quartile of the Social Vulnerability Index. Counties satisfying both conditions are identified as priority counties, while counties satisfying only one condition are retained as comparison groups. This four-part structure matters for management. A county with low experience but lower vulnerability raises a different question from a county with high vulnerability but acceptable experience, and both differ from a county in which low experience and high vulnerability coincide. The latter group is not simply the lowest-scoring group; it represents the intersection of an observed quality concern and a structurally difficult operating environment.

The three maps should therefore be read as a sequence. Figure~\ref{fig:rating_map} shows where overall caregiver experience is stronger or weaker; Figure~\ref{fig:timely_map} tests whether a related operational domain exhibits comparable geographic variation; and Figure~\ref{fig:priority_map} identifies where weak observed experience overlaps with high social vulnerability. Together, they convert fragmented provider reports into an interpretable national monitoring system. They also motivate the regression analysis that follows. The maps establish that geographic variation exists and identify places requiring attention, whereas the adjusted models assess whether provider ownership and county operating context remain associated with caregiver experience after observable differences are considered. This separation between geographic screening and adjusted explanation is essential: the figures support priority setting, but causal attribution and provider accountability require the multivariable evidence presented in Section~\ref{subsec:main_regression_results} and the subsequent validation analyses.

\subsection{Main regression results}
\label{subsec:main_regression_results}

Table~\ref{tab:main_regression_rating} reports the main blockwise regression models for the overall family caregiver rating top-box score. All columns use the same complete-case regression sample of 2,928 providers across 1,078 counties, with standard errors clustered by county FIPS. The specification begins with provider organizational and case-mix variables and sequentially adds rurality, social vulnerability, community health burden, AHRF workforce/resource proxies, and CMS region fixed effects. This structure shows whether county operating context contributes explanatory information beyond provider-level characteristics and makes the adjustment sequence visible to reviewers and decision makers.

The blockwise results provide strong empirical support for the provider--county framework. Adjusted $R^2$ rises from 0.077 in the provider-only model to 0.202 in the preferred model with county context and CMS region fixed effects. Rurality adds 0.037 to adjusted explanatory fit, the community health-burden block adds 0.048, the AHRF workforce and health-resource block adds 0.022, and regional fixed effects add 0.017. County operating context therefore contributes substantial explanatory information beyond provider organization alone.

For-profit ownership is associated with lower overall caregiver rating in every blockwise specification. In the preferred model, for-profit hospices score 3.832 percentage points lower than nonprofit hospices after adjustment for provider characteristics, rurality, social vulnerability, community health burden, county workforce and health-resource context, and CMS region fixed effects ($p<0.01$). Government, other, or unknown ownership is also negatively associated with caregiver rating relative to the nonprofit reference category. The stability of the for-profit estimate across progressively adjusted models shows that the observed ownership difference is not explained by the measured county conditions included in the analysis.

Higher county social vulnerability is associated with lower overall caregiver rating in the preferred model. The nonmetro coefficient becomes small and statistically insignificant after health burden, workforce and health-resource measures, and regional fixed effects are added. This attenuation shows that a simple rural/nonmetro contrast does not capture the broader operating context relevant to hospice caregiver experience.

The PLACES health-burden index has a positive adjusted coefficient. This result is not evidence that poorer population health improves hospice quality. It is an adjusted contextual association that emerges after simultaneous control for vulnerability, rurality, workforce and health-resource measures, and region. Its sign may reflect the joint structure of hospice market maturity, referral patterns, service availability, public-reporting selection, and other correlated county characteristics that are not fully observed in the public data. The coefficient is therefore treated as a contextual adjustment term rather than a direct policy target. AHRF coefficients are interpreted in the same disciplined way: they describe county workforce and health-resource context, not staffing within an individual hospice agency.

\begin{table}[!htbp]
\centering
\caption{Main regression models for overall family caregiver rating}
\label{tab:main_regression_rating}
\tiny
\setlength{\tabcolsep}{1.5pt}
\renewcommand{\arraystretch}{1.02}
\begin{tabular}{@{}L{0.255\textwidth}*{6}{C{0.105\textwidth}}@{}}
\toprule
Variable & M1 & M2 & M3 & M4 & M5 & M6 \\
\midrule
For-profit ownership & -3.950*** & -3.601*** & -3.500*** & -3.762*** & -3.571*** & -3.832*** \\
 & (0.348) & (0.319) & (0.308) & (0.291) & (0.276) & (0.273) \\
Government/other/unknown ownership & -1.832*** & -1.732*** & -1.674*** & -1.897*** & -1.771*** & -1.992*** \\
 & (0.378) & (0.366) & (0.366) & (0.364) & (0.362) & (0.357) \\
Log average daily census & -0.770*** & -0.387*** & -0.364** & -0.351** & -0.190 & -0.122 \\
 & (0.141) & (0.148) & (0.152) & (0.145) & (0.152) & (0.157) \\
Care provided in home (\%) & 0.008 & 0.012 & 0.014 & 0.022** & 0.023** & 0.023** \\
 & (0.011) & (0.010) & (0.010) & (0.009) & (0.009) & (0.009) \\
Care provided in assisted living (\%) & 0.001 & 0.016 & 0.015 & 0.036*** & 0.041*** & 0.053*** \\
 & (0.010) & (0.010) & (0.010) & (0.010) & (0.010) & (0.010) \\
Nonmetro county &  & 3.470*** & 3.465*** & 1.614*** & 0.128 & 0.087 \\
 &  & (0.362) & (0.358) & (0.325) & (0.419) & (0.416) \\
SVI overall percentile &  &  & -0.807 & -5.168*** & -2.026*** & -3.187*** \\
 &  &  & (0.707) & (0.862) & (0.770) & (1.011) \\
PLACES health-burden index &  &  &  & 2.314*** & 1.400*** & 1.698*** \\
 &  &  &  & (0.257) & (0.246) & (0.285) \\
Log APRN availability &  &  &  &  & -0.630*** & -0.518** \\
 &  &  &  &  & (0.216) & (0.223) \\
Primary-care shortage context &  &  &  &  & 0.141 & -0.003 \\
 &  &  &  &  & (0.272) & (0.277) \\
Mental-health shortage context &  &  &  &  & -0.773*** & -0.616** \\
 &  &  &  &  & (0.270) & (0.276) \\
Log health-system infrastructure capacity &  &  &  &  & -0.430 & -0.687** \\
 &  &  &  &  & (0.333) & (0.345) \\
CMS region fixed effects & No & No & No & No & No & Yes \\
Observations & 2928 & 2928 & 2928 & 2928 & 2928 & 2928 \\
Adjusted $R^2$ & 0.077 & 0.114 & 0.115 & 0.163 & 0.185 & 0.202 \\
Incremental $\Delta R^2$ &  & 0.037 & 0.001 & 0.048 & 0.022 & 0.017 \\
\bottomrule
\end{tabular}
\vspace{0.5em}
\begin{minipage}{0.95\textwidth}
\footnotesize\emph{Note.} The dependent variable is the provider-level CAHPS Hospice overall caregiver rating top-box percentage. M1 includes provider organizational controls; M2 adds rurality; M3 adds social vulnerability; M4 adds community health burden; M5 adds workforce and health-resource capacity; M6 adds CMS region fixed effects. Standard errors are reported in parentheses and are county-clustered when feasible. Significance levels: * $p<0.10$, ** $p<0.05$, *** $p<0.01$.
\end{minipage}
\end{table}

\begin{figure}[!tbp]
\centering
\includegraphics[width=0.78\textwidth]{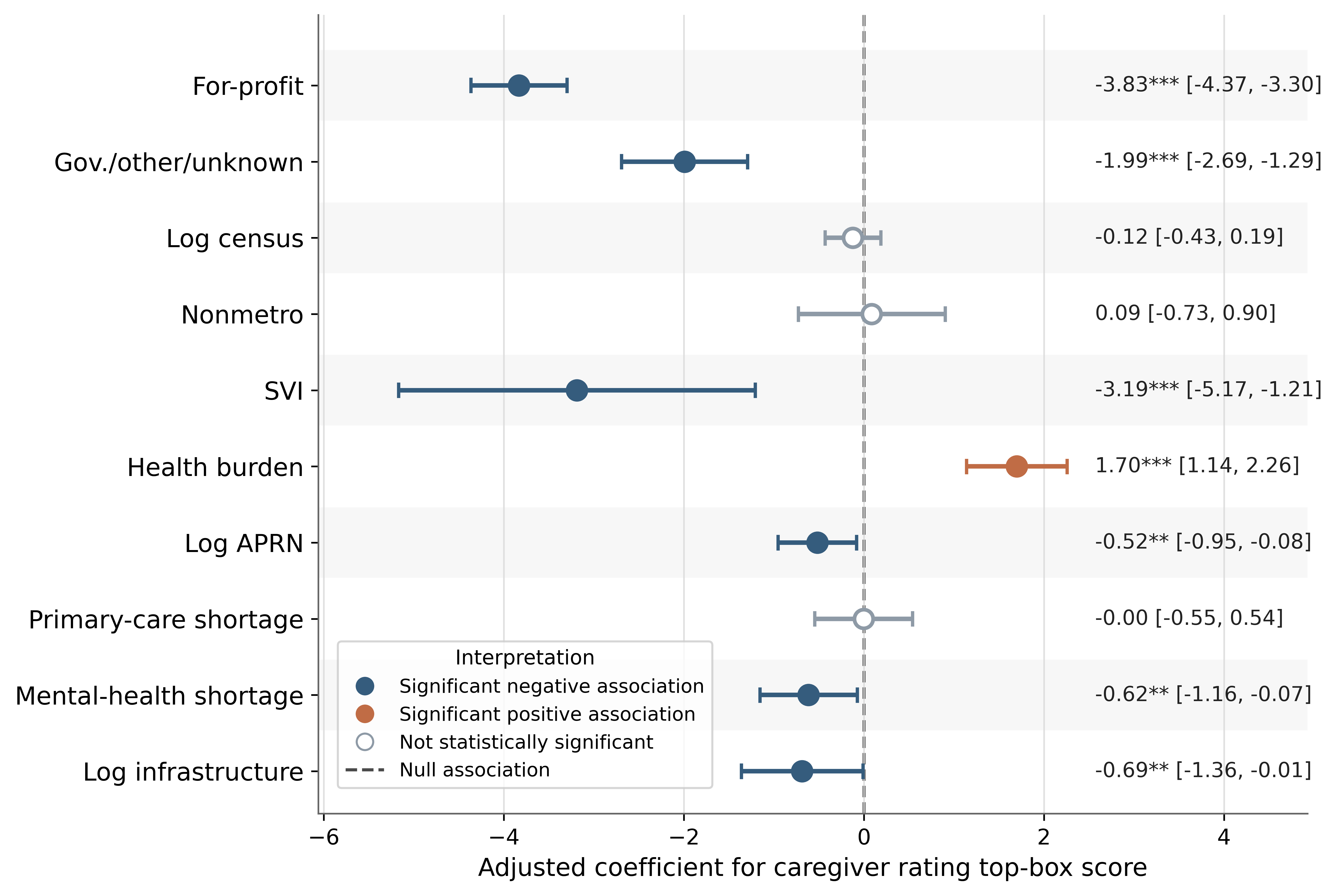}
\caption{Adjusted associations with overall family caregiver rating. Points show coefficient estimates from the preferred model, and horizontal bars indicate 95\% confidence intervals. Filled markers distinguish statistically significant negative and positive associations at the 5\% level; open markers indicate associations that are not statistically significant.}
\label{fig:adjusted_associations_rating}
\end{figure}

Figure~\ref{fig:adjusted_associations_rating} provides a visual summary of the preferred caregiver-rating model. It is positioned immediately after Table~\ref{tab:main_regression_rating} so that reviewers can compare the full regression table with a compact coefficient display. The figure highlights the two most stable negative associations, for-profit ownership and county social vulnerability, while also showing the positive contextual association for the health-burden index and the mixed AHRF proxy results. The display reinforces the paper's central interpretation: caregiver experience should be read as a provider-level outcome shaped jointly by organizational form and local operating conditions.

\subsection{Secondary outcomes and sensitivity analyses}
\label{subsec:secondary_sensitivity}

Table~\ref{tab:secondary_outcomes} reports secondary CAHPS outcome models for timely care, recommendation, team communication, symptom support, respect, and emotional/spiritual support. These models use the same 2,928-provider common sample as the main preferred model, allowing direct comparison across outcome domains. The secondary-outcome table is intentionally placed after the main regression table and Figure~\ref{fig:adjusted_associations_rating}, because it tests whether the main caregiver-rating findings generalize across the broader CAHPS Hospice caregiver-experience profile rather than reflecting a single global rating item.

The ownership result is not confined to one global rating item. For-profit ownership is negatively associated with all six secondary CAHPS Hospice outcomes: timely care, willingness to recommend, team communication, symptom support, respectful treatment, and emotional or spiritual support. The adjusted differences range from 1.524 percentage points for emotional or spiritual support to 4.621 percentage points for willingness to recommend, and all six estimates are statistically significant. This cross-domain consistency is one of the study's strongest findings. It shows that the ownership association extends across multiple components of service delivery rather than reflecting only a general impression captured by the overall rating.

County social vulnerability is also negatively associated with every secondary caregiver-experience outcome. The pattern spans responsiveness, communication, symptom support, respect, caregiver preparation, and recommendation, indicating that the vulnerability association is broad rather than domain-specific. The nonmetro indicator is not independently associated with any secondary outcome after full adjustment. Mental-health workforce-shortage context is negatively associated with timely care, recommendation, team communication, symptom support, and respect, but not with emotional or spiritual support. Because this measure characterizes county workforce context rather than hospice-level staffing, it should be interpreted as evidence that local mental-health resource constraints accompany weaker caregiver experience in several domains, not as a direct estimate of hospice staffing effects.

\begin{table}[!htbp]
\centering
\caption{Secondary CAHPS Hospice caregiver-experience outcome regressions}
\label{tab:secondary_outcomes}
\tiny
\setlength{\tabcolsep}{1.3pt}
\renewcommand{\arraystretch}{1.02}
\begin{tabular}{@{}L{0.245\textwidth}*{6}{C{0.112\textwidth}}@{}}
\toprule
Variable & Timely care & Recommend & Team comm. & Symptoms & Respect & Emotional support \\
\midrule
For-profit ownership & -3.056*** & -4.621*** & -2.432*** & -2.518*** & -1.720*** & -1.524*** \\
 & (0.299) & (0.295) & (0.229) & (0.256) & (0.158) & (0.153) \\
Government/other/unknown ownership & -1.688*** & -2.694*** & -1.313*** & -1.570*** & -1.034*** & -0.811*** \\
 & (0.400) & (0.384) & (0.307) & (0.360) & (0.215) & (0.213) \\
Log average daily census & -0.670*** & 0.285* & -0.374*** & -0.495*** & -0.288*** & -0.023 \\
 & (0.154) & (0.162) & (0.126) & (0.150) & (0.095) & (0.102) \\
Care provided in home (\%) & 0.006 & 0.036*** & 0.048*** & 0.048*** & 0.014*** & -0.005 \\
 & (0.009) & (0.010) & (0.007) & (0.009) & (0.005) & (0.006) \\
Care provided in assisted living (\%) & 0.022** & 0.058*** & 0.031*** & 0.044*** & 0.034*** & 0.028*** \\
 & (0.011) & (0.011) & (0.008) & (0.009) & (0.006) & (0.006) \\
Nonmetro county & 0.268 & 0.351 & -0.057 & -0.166 & 0.138 & 0.299 \\
 & (0.469) & (0.436) & (0.330) & (0.385) & (0.252) & (0.220) \\
SVI overall percentile & -4.546*** & -3.613*** & -2.747*** & -2.489*** & -2.609*** & -2.204*** \\
 & (1.108) & (1.035) & (0.833) & (0.922) & (0.592) & (0.531) \\
PLACES health-burden index & 2.616*** & 1.759*** & 1.483*** & 1.767*** & 1.080*** & 0.719*** \\
 & (0.320) & (0.304) & (0.243) & (0.264) & (0.183) & (0.172) \\
Log APRN availability & -0.404* & -0.786*** & -0.411** & -0.511*** & -0.277** & -0.252*** \\
 & (0.237) & (0.227) & (0.165) & (0.166) & (0.128) & (0.092) \\
Primary-care shortage context & 0.034 & -0.059 & -0.296 & -0.235 & -0.048 & 0.049 \\
 & (0.289) & (0.300) & (0.211) & (0.270) & (0.154) & (0.153) \\
Mental-health shortage context & -0.671** & -0.595** & -0.527** & -0.874*** & -0.299* & -0.046 \\
 & (0.291) & (0.295) & (0.216) & (0.268) & (0.158) & (0.169) \\
Log health-system infrastructure capacity & -0.627* & -0.633** & -0.451* & -0.333 & -0.328 & -0.542** \\
 & (0.360) & (0.316) & (0.270) & (0.257) & (0.216) & (0.216) \\
CMS region fixed effects & Yes & Yes & Yes & Yes & Yes & Yes \\
Observations & 2928 & 2928 & 2928 & 2928 & 2928 & 2928 \\
Adjusted $R^2$ & 0.240 & 0.238 & 0.222 & 0.200 & 0.189 & 0.178 \\
\bottomrule
\end{tabular}
\vspace{0.5em}
\begin{minipage}{0.95\textwidth}
\footnotesize\emph{Note.} Each column reports the preferred adjusted model for a secondary CAHPS Hospice caregiver-experience top-box outcome. The covariate set matches the preferred model in Table~\ref{tab:main_regression_rating}: provider organizational controls, rurality, social vulnerability, community health burden, AHRF workforce and health-resource capacity proxies, and CMS region fixed effects. Standard errors are reported in parentheses and are county-clustered when feasible. AHRF variables are county-level workforce and health-resource environment proxies and should not be interpreted as hospice-specific staffing levels. Significance levels: * $p<0.10$, ** $p<0.05$, *** $p<0.01$.
\end{minipage}
\end{table}

Table~\ref{tab:sensitivity_analyses} summarizes the sensitivity analyses. It is placed immediately after the secondary-outcome table because it addresses the main robustness questions a reviewer is likely to ask: whether the findings depend on regional fixed-effect choice, the PLACES health-burden index, the AHRF block, sparse-county observations, star-rating sample restrictions, or the use of overall rating rather than timely care as the primary outcome. The ownership result is highly stable: for-profit ownership remains negative and statistically significant in all sensitivity specifications. The social-vulnerability result is also broadly stable, remaining negative in seven of eight specifications and statistically significant in seven specifications. It attenuates when the PLACES health-burden index is excluded, indicating shared contextual variation between social vulnerability and community health burden. The health-burden index remains positive and statistically significant whenever included, reinforcing the need for cautious contextual rather than causal interpretation.

\begin{table}[tbp]
\centering
\caption{Sensitivity analyses for the main caregiver-experience findings}
\label{tab:sensitivity_analyses}
\fontsize{6.4}{7.1}\selectfont
\setlength{\tabcolsep}{1.5pt}
\renewcommand{\arraystretch}{1.02}
\begin{tabular}{p{0.035\textwidth}p{0.205\textwidth}p{0.085\textwidth}p{0.052\textwidth}p{0.13\textwidth}p{0.13\textwidth}p{0.13\textwidth}p{0.13\textwidth}}
\toprule
Spec. & Sensitivity analysis & N (counties) & Adj. $R^2$ & For-profit ownership & SVI overall percentile & PLACES health-burden index & Mental-health shortage context \\
\midrule
S0 & Preferred model & 2928 (1078) & 0.202 & -3.832*** (0.273) & -3.187*** (1.011) & 1.698*** (0.285) & -0.616** (0.276) \\
S1 & State fixed effects & 2928 (1078) & 0.218 & -4.135*** (0.271) & -2.877** (1.305) & 1.433*** (0.469) & -0.499* (0.284) \\
S2 & No health-burden index & 3098 (1141) & 0.178 & -3.519*** (0.266) & 0.560 (0.746) & Not included & -0.470* (0.285) \\
S3 & No AHRF block & 2928 (1078) & 0.182 & -4.036*** (0.279) & -6.521*** (0.957) & 2.666*** (0.282) & Not included \\
S4 & Reduced AHRF block & 2928 (1078) & 0.187 & -4.038*** (0.278) & -5.873*** (0.967) & 2.569*** (0.273) & -1.023*** (0.265) \\
S5 & Counties with $\geq$2 rating providers & 2325 (475) & 0.168 & -4.027*** (0.328) & -3.860*** (1.388) & 2.128*** (0.382) & -0.764** (0.344) \\
S6 & Star-rating available sample & 1904 (875) & 0.239 & -3.743*** (0.292) & -3.399*** (0.861) & 1.757*** (0.271) & -0.578** (0.262) \\
S7 & Timely care outcome & 2928 (1078) & 0.240 & -3.056*** (0.299) & -4.546*** (1.108) & 2.616*** (0.320) & -0.671** (0.291) \\
\bottomrule
\end{tabular}
\vspace{0.5em}
\begin{minipage}{0.95\textwidth}
\footnotesize\emph{Note.} Entries are adjusted coefficients with county-clustered standard errors in parentheses. S0 is the preferred caregiver-rating model. S1 replaces CMS region fixed effects with state fixed effects. S2 excludes the PLACES health-burden index. S3 excludes the AHRF block. S4 uses a reduced AHRF block. S5 restricts the sample to counties with at least two CAHPS-rating providers. S6 restricts to providers with nonmissing summary star rating. S7 uses timely care as an alternative primary outcome. Significance levels: * $p<0.10$, ** $p<0.05$, *** $p<0.01$.
\end{minipage}
\end{table}

The evidence across Tables~\ref{tab:main_regression_rating}--\ref{tab:sensitivity_analyses} closes the empirical loop of the provider--county assessment framework. The descriptive maps identify where caregiver-experience differences and high vulnerability appear geographically, while the main regression table shows how ownership and county context relate to caregiver rating after adjustment. The secondary-outcome models indicate that the same organizational and vulnerability signals extend beyond a single global rating item, and the sensitivity analyses show that the principal ownership and vulnerability findings are not driven by one fixed-effect structure, one sample restriction, or one contextual block. In applied HCMS terms, this is the validation layer of the study: it does not convert the analysis into a causal design, but it shows that the framework yields stable and interpretable performance information for context-aware hospice monitoring.

\FloatBarrier

\section{Practical and Policy Implications}
\label{sec:implications}
% ============================================================

The study resolves the performance-assessment problem posed in the Introduction. Hospice caregiver experience cannot be understood adequately from provider characteristics alone, but neither can observed differences be attributed solely to the communities in which agencies operate. Adding county context increased adjusted \(R^2\) from 0.077 in the provider-only model to 0.202 in the preferred model. This gain shows that rurality, social vulnerability, community health burden, workforce and health-resource conditions, and regional structure contribute meaningful information to national provider assessment. This pattern is consistent with context-aware health-care performance research showing that organizational results must be interpreted in relation to local need, resource availability, and feasible operating conditions \cite{Merelie2025DeprivationDEA,Camanho2006Malmquist}. Yet contextual adjustment did not remove the ownership difference. The evidence therefore supports a provider--county model of accountability: the hospice remains responsible for the care it organizes, while comparison and intervention take account of the operating conditions surrounding that care.

\subsection{Ownership, governance, and quality accountability}

The most stable empirical result concerns ownership. In the preferred model, for-profit hospices had overall caregiver ratings 3.832 percentage points below nonprofit hospices after extensive provider and county adjustment. The same association appeared in every blockwise specification, all six secondary CAHPS Hospice domains, and all eight sensitivity specifications. Across the secondary outcomes, the adjusted difference ranged from 1.524 percentage points for emotional or spiritual support to 4.621 points for willingness to recommend; timely care was 3.056 points lower, team communication 2.432 points lower, symptom support 2.518 points lower, and respectful treatment 1.720 points lower.

This breadth is more important than any single coefficient. It indicates that the ownership association is not limited to a general rating but extends across responsiveness, communication, symptom support, respect, caregiver preparation, and recommendation. The analysis does not identify a causal mechanism, and the broad CMS category does not distinguish private-equity or publicly traded ownership from other for-profit forms. It nevertheless establishes that the observed disadvantage persists after adjustment for agency scale, care setting, patient composition, rurality, vulnerability, health burden, resource context, and region. Ownership should therefore remain a central stratification variable in hospice quality surveillance and governance review. This conclusion complements evidence that private-equity and publicly traded ownership can be associated with lower caregiver-reported quality, while broader for-profit entry may also expand access and alter Medicare spending through different pathways \cite{Aldridge2024HospicePEOpacity,Soltoff2024JAMAHospiceOwnership,Gruber2025DyingOrLying}. For managers and regulators, the appropriate next question is not whether all for-profit agencies are alike, but which organizational practices---including visit organization, after-hours response, interdisciplinary coordination, staffing priorities, caregiver preparation, and ownership transparency---may explain the persistent cross-domain difference.

\subsection{Equity, geographic priority, and local capacity}

County social vulnerability identifies a second and distinct source of disadvantage. The result accords with established access and neighborhood-context frameworks in which socioeconomic disadvantage, transportation barriers, social support, and local resource availability shape both care needs and the feasibility of service delivery \cite{Flanagan2011SVI,Andersen1995BehavioralModel,KindBuckingham2018NeighborhoodAtlas}. In the preferred model, the Social Vulnerability Index coefficient was -3.187 percentage points, and negative associations appeared across all six secondary outcomes. The geographic analysis adds practical meaning to this result by identifying counties in which low caregiver ratings and high vulnerability coincide. These priority counties are not a league table of provider quality. They are places where a weak observed service signal and a difficult social environment occur together and where uniform national intervention is least likely to be adequate.

The appropriate response is differentiated. When adjusted provider performance is weak but structural constraints are less severe, organizational review and quality-improvement support should take priority. Where vulnerability and local resource constraints are substantial but adjusted performance is not unusually poor, area-level investments in transportation, workforce, behavioral-health coordination, and caregiver support may be more relevant. Where both conditions coexist, provider-level and area-level interventions should be combined. This is the practical value of linking the priority map to adjusted provider models rather than relying on either raw rankings or geographic description alone.

The mental-health workforce result reinforces this interpretation. Mental-health shortage context was associated with a 0.616-percentage-point reduction in the overall rating and with lower timely care, recommendation, communication, symptom support, and respectful treatment. These estimates describe county resource conditions rather than agency staffing. They nonetheless show that hospice quality management cannot be isolated from the behavioral-health and community-service networks on which home-based end-of-life care depends. For provider leaders, quality-improvement planning should therefore examine both internal workflow and the external referral and support system. This interpretation is consistent with evidence that staffing, workload, and deployable capacity affect quality, timeliness, burnout, and cost across health-care settings \cite{Aiken2002NurseStaffing,Cho2023PatientNurseRatios,Sulz2024PatientBasedNurseStaffing}. For policymakers, hospice oversight should be coordinated with local workforce and service-capacity planning, including the geographic and capacity constraints emphasized in health-care assignment and long-term-care planning research \cite{Schafer2023PatientBedAssignment,Khalili2025LTCCapacityPlanning}.

The social-vulnerability finding also requires an exact boundary. Its association was negative in the preferred model, all six secondary outcomes, and most sensitivity specifications, but it attenuated when the community health-burden block was removed. The defensible interpretation is that vulnerability is broadly related to poorer caregiver experience while sharing part of its explanatory structure with correlated county health conditions. This qualification does not diminish the equity finding; it prevents an overly simple claim that a single county index operates independently of the wider environment in every specification.

\subsection{Implications for national measurement and Medicare stewardship}

The results answer the three concerns raised at the outset and connect the empirical findings to current national measurement and oversight priorities \cite{MedPAC2026Hospice,HHSOIG2026HospiceEligibility}. First, for protection of end-of-life quality, the analysis identifies a stable and multidomain ownership-related disadvantage rather than relying on a single national average or one process indicator. Second, for stewardship of Medicare resources, it provides a reproducible method for directing monitoring, technical assistance, and quality-improvement attention toward patterns that remain visible after adjustment. The study does not estimate improper payment or causal savings, but in a sector serving 1.82 million beneficiaries and accounting for \$28.3 billion in annual Medicare spending, a more discriminating allocation of oversight resources is itself an important management contribution. Third, for accountability in a changing provider market, the findings show that local disadvantage matters but does not explain away the observed ownership difference.

The framework also clarifies what context-aware benchmarking should accomplish. It should not lower expectations for agencies serving disadvantaged communities, nor should it compare all agencies as though their operating environments were interchangeable. Instead, it should identify providers performing below expectations under comparable conditions and providers achieving strong family experience despite substantial constraints. The first group warrants closer examination for remediable weaknesses; the second offers credible peers from which operational practices can be learned.

The broader contribution to health-care management science is an auditable decision structure that joins provider accountability, geographic context, multidomain caregiver experience, and robustness analysis. It follows the central benchmarking principle that performance should be assessed relative to relevant peers and operating conditions rather than through unadjusted ranking alone \cite{Charnes1978DEA,Banker1984DEA,Camanho2006Malmquist,Merelie2025DeprivationDEA}. By preserving the hospice agency as the managerial unit while incorporating the county as an operating context, the study moves beyond both unadjusted ranking and purely geographic description. It provides a defensible basis for peer comparison, targeted quality improvement, and resource planning while maintaining the central principle established by the evidence: community conditions must be considered, but they do not eliminate organizational responsibility.

% ============================================================
\section{Conclusion}
\label{sec:conclusion}
% ============================================================

This study develops a national provider--county framework for assessing
hospice caregiver experience while preserving the hospice agency as the
unit of accountability. Across 2,928 hospices in 1,078 counties,
for-profit ownership was associated with lower overall caregiver ratings
after adjustment for provider characteristics and county operating
context. The same negative association appeared across all six secondary
Consumer Assessment of Healthcare Providers and Systems (CAHPS) Hospice
domains and remained statistically significant in every sensitivity
specification.

County context also materially improved performance assessment. Adjusted
\(R^2\) increased from 0.077 in the provider-only model to 0.202 in the
preferred provider--county model, and higher social vulnerability was
associated with poorer caregiver experience in the preferred model and
across all secondary domains. These findings show that hospice agencies
do not operate under interchangeable local conditions, yet contextual
adjustment does not eliminate the observed ownership difference.

The study therefore contributes a transparent national assessment
structure that separates, as far as public data allow, provider-level
responsibility from county-level operating constraint. By identifying
where weak caregiver experience overlaps with difficult local conditions,
the framework supports context-aware monitoring, peer comparison,
targeted technical assistance, and more defensible allocation of
quality-improvement resources.

The findings represent adjusted associations rather than causal effects
and are most directly applicable to hospices included in the public
reporting sample
\cite{Wooldridge2010EconometricAnalysis,
ImbensRubin2015CausalInference,HernanRobins2020CausalInference}.
County indicators characterize local operating context rather than
agency-level staffing, patient-level complexity, or complete service
areas.

Future research can extend the framework using longitudinal CAHPS data,
provider-level staffing measures, detailed service-area information, and
ownership data that distinguish private-equity and publicly traded
organizations from other for-profit forms.

% ============================================================
\begin{appendices}
% ============================================================
\renewcommand{\theequation}{\Alph{section}\arabic{equation}}
\renewcommand{\thetable}{\Alph{section}\arabic{table}}
\renewcommand{\thefigure}{\Alph{section}\arabic{figure}}
\renewcommand{\theHequation}{\Alph{section}.\arabic{equation}}
\renewcommand{\theHtable}{\Alph{section}.\arabic{table}}
\renewcommand{\theHfigure}{\Alph{section}.\arabic{figure}}

\section{Data linkage audit}
\setcounter{equation}{0}\setcounter{table}{0}\setcounter{figure}{0}

A file-level audit was conducted for all public data sources used in the analysis. The audit recorded row and column counts, key identifiers, encoding checks, and SHA-256 checksums before linkage. Provider files were joined using the Centers for Medicare \& Medicaid Services Certification Number, and county files were joined using five-digit Federal Information Processing Standards codes.

The final provider-level analytic base retained 6,852 hospice records after county-covariate linkage, confirming that no row inflation occurred during merging. The harmonized county-covariate file contained 3,242 records and 3,242 unique county Federal Information Processing Standards codes, with no duplicate county keys. The Area Health Resources Files source contained 3,235 records and 3,235 unique county codes, also with no duplicates. Differences in county counts across sources therefore reflect source-specific geographic coverage rather than many-to-many merge artifacts.

\section{Variable definitions and construction}
\setcounter{equation}{0}\setcounter{table}{0}\setcounter{figure}{0}

Table~\ref{tab:variable_dictionary} summarizes the principal variables used in the provider--county models. Exact source-field names, coding decisions, and reproducible transformations are preserved in the deposited analysis records and code.

\begin{table}[!htbp]
\centering
\caption{Principal analytical variables and construction}
\label{tab:variable_dictionary}
\scriptsize
\setlength{\tabcolsep}{3pt}
\renewcommand{\arraystretch}{0.96}
\begin{tabular}{p{0.18\textwidth}p{0.18\textwidth}p{0.25\textwidth}p{0.25\textwidth}}
\toprule
Construct & Source & Construction & Analytical interpretation \\
\midrule
Overall caregiver rating & CAHPS Hospice Survey & Percentage of respondents rating the hospice 9 or 10 & Primary caregiver-experience outcome \\
Willingness to recommend & CAHPS Hospice Survey & Provider-level top-box percentage & Secondary caregiver-experience outcome \\
Timely care & CAHPS Hospice Survey & Provider-level top-box percentage & Responsiveness and deployable service capacity \\
Team communication & CAHPS Hospice Survey & Provider-level top-box percentage & Clarity and continuity of information transfer \\
Symptom support & CAHPS Hospice Survey & Provider-level top-box percentage & Coordination between assessment and symptom response \\
Respectful treatment & CAHPS Hospice Survey & Provider-level top-box percentage & Respect and dignity experienced by patients and families \\
Emotional or spiritual support & CAHPS Hospice Survey & Provider-level top-box percentage & Support for emotional and spiritual needs \\
Ownership & CMS Hospice General Information & Categorical indicators with nonprofit ownership as the reference group & Provider governance and organizational structure \\
Provider scale and service mix & CMS Hospice Provider Data and utilization files & Provider-level measures including average daily census, care setting, service type, payer mix, and patient composition & Organizational scale and operating profile \\
Rurality & USDA Rural--Urban Continuum Codes & Metropolitan or nonmetropolitan classification derived from county code & Geographic operating context \\
Social vulnerability & CDC/ATSDR Social Vulnerability Index & Overall county percentile rank on a 0--1 scale & County social and socioeconomic vulnerability \\
Community health burden & CDC PLACES & Standardized composite of selected county prevalence measures & County population-health context \\
Advanced practice registered nurse availability & Area Health Resources Files & County workforce-availability measure; transformed as specified in the analysis code & Local clinical workforce environment \\
Primary-care shortage context & Area Health Resources Files & County shortage indicator & Local primary-care resource constraint \\
Mental-health shortage context & Area Health Resources Files & County shortage indicator & Local mental-health workforce constraint \\
Health-system infrastructure capacity & Area Health Resources Files & County health-resource capacity measure & Local service-system infrastructure \\
\bottomrule
\end{tabular}
\end{table}

County workforce and health-resource variables characterize the local operating environment and are not interpreted as direct measures of staffing within an individual hospice. Similarly, county health-burden measures describe population context rather than the clinical complexity of a provider's own patient panel.

\section{County FIPS assignment and linkage rules}
\setcounter{equation}{0}\setcounter{table}{0}\setcounter{figure}{0}

County Federal Information Processing Standards codes were assigned using provider state and county names matched to the U.S. Department of Agriculture Rural--Urban Continuum Codes county identifiers. District of Columbia providers with unresolved county codes were assigned code 11001. Providers without a resolved county code were retained in provider-level files but excluded from models requiring county covariates.

County-level public sources differed slightly in geographic coverage. The harmonized county table therefore used the union of valid county codes available across sources while preserving one unique record per county. Unmatched and source-specific county-equivalent records were documented in the linkage audit. This procedure preserved valid observations without introducing duplicate county keys or provider-row inflation.

\end{appendices}

% ============================================================
% Ethics approval
% ============================================================
\section*{Ethics approval}

This study used publicly available secondary data, including provider-level Centers for Medicare \& Medicaid Services Hospice and Consumer Assessment of Healthcare Providers and Systems Hospice Survey data and county-level public data sources. The authors did not collect primary survey data, contact human participants, or use identifiable individual-level records, patient-level data, or protected health information. Ethics approval was not required because the analysis was limited to publicly available, de-identified secondary data.

% ============================================================
% Data availability
% ============================================================
\section*{Data Availability}

The reproducibility materials for this study are deposited in Zenodo under the title \textit{Reproducibility Package for ``Geographic Disparities in Hospice Quality and Family Caregiver Experience: Ownership, Social Vulnerability, and County Operating Context}. The version-specific DOI is \url{https://doi.org/10.5281/zenodo.21217531}, and the concept DOI for all versions is \url{https://doi.org/10.5281/zenodo.21217530}.

The study is based on publicly available secondary data downloaded from the original public data providers and linked for analysis. These sources include provider-level Centers for Medicare \& Medicaid Services Hospice and Consumer Assessment of Healthcare Providers and Systems Hospice Survey data, the Health Resources and Services Administration Area Health Resources Files, the United States Department of Agriculture Rural--Urban Continuum Codes, the Centers for Disease Control and Prevention/Agency for Toxic Substances and Disease Registry Social Vulnerability Index, the Centers for Disease Control and Prevention Population Level Analysis and Community Estimates program, and U.S. Census county boundary files.

The deposited package contains the source-data manifest, provider- and county-level linkage documentation, variable-construction records, audit summaries, derived analysis files, model specifications, analysis code, and the workflow used to generate the reported tables and figures. Raw public source files are not redistributed because they remain available from their original public providers.

No restricted patient-level data, identifiable individual-level data, or protected health information are included in the study or in the reproducibility package.

% ============================================================
% References
% ============================================================
\bibliography{HCMS_references}

\end{document}